\documentstyle[12pt,epsf]{article}
\begin{document}
\baselineskip 16pt plus 2pt minus 2pt

\begin{titlepage}

\par
\topmargin=-1cm      

{ \small

\noindent{FZJ-IKP(TH)-02-10} \hfill{NT@UW-01-031}

}

\vspace{20.0pt}

\begin{centering}

{\large\bf{The S-Wave Pion-Nucleon Scattering Lengths\\[0.2em] 
from Pionic Atoms using Effective Field Theory}}

\vspace{40.0pt}
{{\bf S.R.~Beane}$^1$,
{\bf V.~Bernard}$^2$,
{\bf E.~Epelbaum}$^3$,\\
{\bf Ulf-G.~Mei{\ss}ner}$^{4,5}$ and
{\bf D.R.~Phillips}$^6$}\\
\vspace{15.0pt}
{\sl $^{1}$Department of Physics, University of Washington,
Seattle, WA 98195, USA}\\
{\it E-mail: sbeane@phys.washington.edu}\\  
\vspace{14.0pt}
{\sl $^{2}$Laboratoire de Physique Th\'eorique \\
Universit\'e Louis Pasteur, F-67084 Strasbourg, France} \\
{\it E-mail: bernard@lpt6.u-strasbg.fr}\\
\vspace{14.0pt}
$^{3}${\sl Ruhr-Universit\"at Bochum, Institut f{\"u}r
  Theoretische Physik II,\\ D-44870 Bochum, Germany}\\
{\it E-mail: evgeni.epelbaum@tp2.ruhr-uni-bochum.de }\\
\vspace{14.0pt}
{\sl $^{4}$Forschungszentrum J\"ulich, Institut f\"ur Kernphysik (Th), D-52425 J\"ulich,
    Germany}\\
{\it E-mail: u.meissner@fz-juelich.de }\\
\vspace{14.0pt}
{\sl $^{5}$Karl-Franzens-Universit\"at Graz, Institut f\"ur Theoretische Physik, A-8010 Graz, Austria}\\
\vspace{14.0pt}
{\sl $^{6}$Department of Physics and Astronomy, Ohio
      University, Athens, OH 45701, USA}
{\it E-mail: phillips@phy.ohiou.edu }\\

\vspace{7.0pt}
\end{centering}
\vspace{12.0pt}

\begin{abstract}
\noindent 

The pion-deuteron scattering length is computed to next-to-next-to-leading
order in baryon chiral perturbation theory.  A modified power-counting is then
formulated which properly accounts for infrared enhancements engendered by the
large size of the deuteron, as compared to the pion Compton wavelength. We use
the precise experimental value of the real part of the pion-deuteron
scattering length determined from the decay of pionic deuterium, together with
constraints on pion-nucleon scattering lengths from the decay of pionic
hydrogen, to extract the isovector and isoscalar S-wave pion-nucleon scattering
lengths, $a^-$ and $a^+$, respectively. We find $a^{-}=[0.0918 (\pm 0.0013)\;] {M_\pi^{-1}}$ and
$a^{+}=[-0.0034 (\pm 0.0007)\;] {M_\pi^{-1}}$.

\end{abstract}

\vspace*{5pt}
\begin{center}
\end{center}
\vfill

\end{titlepage}

\section{Introduction}
\label{sec:intro}

\noindent The pion-nucleon ($\pi$-N) scattering lengths are quantities of
fundamental importance in hadronic physics. They provide an
important test of QCD symmetries and the pattern of their breaking.
They also provide a crucial constraint on the $\pi$-N
interaction and, as such, have an impact on our understanding of
nucleon-nucleon (NN) scattering, the three-nucleon force, and
pion-nucleus scattering.

In the limit of exact isospin symmetry the threshold $\pi$-N amplitude
may be written as
\begin{equation}
T_{\pi N}^{ba}=4 \pi (1 + \mu) \left[\delta^{ba} a^+ + i \epsilon^{b
a c} \tau^c a^-\right] \ \ ,
\end{equation}
where $\mu \equiv M_\pi/m\simeq 1/7$ is the genuine small threshold 
parameter, and $a^+$ and $a^-$ are the isoscalar and
isovector S-wave scattering lengths, respectively. 
While these quantities cannot be measured directly in $\pi$-N scattering
experiments, the well-known effective field theory (EFT), 
baryon chiral perturbation theory ($\chi$PT), offers an ideal tool 
for extrapolating the $\pi$-N amplitude to threshold.
Working up to fourth order in the dual momentum and pion-mass
expansion, the values shown in Table~\ref{table-pinas} have been obtained from various
analyses of $\pi$-N data~\cite{moj,Fettes:1998ud,Fettes:2000xg,Buttiker:2000ap}.
(Note that there are no fourth-order corrections to $a^-$~\cite{Bernard:1995pa}. 
The difference between the $O(p^3)$ and $O(p^4)$ values in 
Table~\ref{table-pinas} arises from a refitting of low-energy constants at
lower orders, as discussed in Ref.~\cite{Fettes:2000xg}.)

\begin{table}[bht]
\begin{center}
\begin{tabular}{|r|r|r|r|r|}
    \hline
          & $O(p)$  &   $O(p^2)$  &  $O(p^3)$ 
          & $O(p^4)$  \\
    \hline\hline
           Fit 1      & 0.0  & 0.0046  & $-$0.0100    & $-$0.0096 \\
$a^+ \quad$ Fit 2    & 0.0  & 0.0024  &    0.0049  &    0.0045 \\
           Fit 3      & 0.0  & 0.0101  &    0.0014  &    0.0027 \\ 
\hline
           Fit 1      & 0.0790 & 0.0790  &  0.0905  &    0.0903 \\
$a^- \quad$ Fit 2     & 0.0790 & 0.0790  &  0.0772  &    0.0771 \\
           Fit 3      & 0.0790 & 0.0790  &  0.0870  &    0.0867 \\
\hline\hline
  \end{tabular}
  \caption{Convergence of the S--wave scattering lengths for $\chi$PT
   fits to the Karlsruhe~\cite{KA85}, Matsinos~\cite{Ma98}, and VPI~\cite{EM98}
    phase-shift analyses of $\pi$-N data. 
   $O(p^n)$ means that all terms up-to-and-including
   order $n$ were included in the $\chi$PT fit. 
   Units are $M_\pi^{-1}$. Table taken from Ref.~\cite{Fettes:2000xg}.
    \label{table-pinas}}
\end{center}\end{table}

An independent approach to the $\pi$-N scattering lengths involves
analyzing pionic-atom level shifts and widths. In the Coulombic
$\pi^-$-p system, the strong-interaction shift in the energy 
can be used to infer a value for $a^-$. 
The recent Neuchatel-PSI-ETHZ (NPE) experiment~\cite{Schroder:uq} finds:
\begin{equation}
a^-=[0.0905(\pm 0.0042)] M_\pi^{-1} \ \ .
\label{eq:NPEa-}
\end{equation}
Note that the error here is
significantly smaller than the spread of values obtained from the
analyses of $\pi$-N scattering (see Table~\ref{table-pinas}). 
Only the lifetime
(or, equivalently, the width) of pionic hydrogen is sensitive to
$a^+$. The NPE experiment~\cite{Schroder:uq} finds the value:
\begin{equation}
a^+=[-0.0022(\pm 0.0043)] M_\pi^{-1}\ \ .
\end{equation}

Thus, neither $\pi$-N scattering nor the $\pi^-$-p atom provide a
strong constraint on $a^+$. The isoscalar scattering length may well
be probed more directly in the $\pi^-$-deuterium atom. The NPE measurement~\cite{Hauser:1998yd}
of the pionic-deuterium atomic-level shift yields:
\begin{equation}
a_{\pi d}=[-0.0261 (\pm 0.0005)\;+\; i\; 0.0063 (\pm 0.0007)\;] {M_\pi^{-1}} 
\label{eq:pionicdeut}
\end{equation}
for the pion-deuteron ($\pi$-d) scattering length---a measurement that is
remarkably accurate given the usual level of precision in
hadronic-physics experiments~\footnote{For a general introduction to
the $\pi$-d scattering length and its significance see
Ref.~\cite{eric}.}.  In order to precisely relate this number to
$a^+$ an EFT of threshold $\pi$-d scattering is required. 

There has been remarkable recent progress in developing the 
EFTs relevant to multi-nucleon systems~\cite{Epelbaum:2001ue,Ioffe}.
Among the advantages of this formalism is a quantitative method for estimating
theoretical errors and a unified field-theoretic treatment of processes involving
different numbers of nucleons. The non-perturbative effects responsible for
deuteron binding are accounted for in the EFT power-counting, with a meaningful
quantification of the theoretical error. The multi-nucleon EFT relevant to
momentum transfers of order the pion mass is in most cases a straightforward 
generalization of single-nucleon baryon $\chi$PT~\cite{Weinberg:1990rz,Weinberg:1991um,Beane:2001bc}. 
NN phase shifts and deuteron properties have now been computed to
next-to-next-to-leading order (NNLO=$O(p^3)$ for the NN potential,
where $p$ denotes a small momentum/pion mass) in this
expansion~\cite{Ordonez:1996rz,Epelbaoum:1998ka,Epelbaum:2000dj}. Processes
with external pions and photons have also been extensively studied~\cite{Ioffe}.

Consider $\pi$-d scattering at
threshold~\cite{wein1,Beane:1998y}. Two-pion-four-nucleon operators
which contribute to $\pi$-d scattering enter at high orders in the EFT
expansion and are therefore highly suppressed~\cite{wein1}. That these
contributions are very small is crucial to the predictive power of the EFT, as
the coefficients which determine the strength of these operators are unknown,
and need therefore be determined from a separate nuclear observable, or, at
some time in the distant futurity, from lattice QCD~\cite{Beane:2001bc}. This
suppression of local operators has an associated benefit: the $\pi$-d
scattering length is sensitive to low-energy constants which contribute to the
S-wave $\pi$-N scattering lengths, $a^-$ and $a^+$. 

While baryon $\chi$PT provides a rigorous theory of low-energy $\pi$-d scattering,
there is an important kinematical subtlety in the threshold region; this leads 
to a new EFT~\cite{BeSa} which distinguishes between the low scales $M_\pi$ and $\gamma$,
where $\gamma$ is the deuteron binding momentum. This
straightforward generalization of baryon $\chi$PT turns out to be an extremely
useful tool in developing a precise theoretical relation between the $\pi$-d
and $\pi$-N S-wave scattering lengths.

Some additional remarks are in order. First, we should mention
that the EFTs used in this paper can be applied to scattering and bound-state
observables. For the direct extraction of the S-wave scattering lengths from
pionic atoms one can also use the bound-state EFT as applied, for example, to $\pi^-$-p
atoms in Ref.~\cite{Lyubovitskij:2000kk}.  Since that approach is designed to deal
with the pionic-hydrogen system, it is expected to give better control of the
theoretical errors.  However, for the case of pionic deuterium one faces a
genuine three-body problem, which so far has not been formulated in the
bound-state EFT, whereas it is straightforward to deal with in our approach.
Second, throughout this work, we neglect isospin violation. 
Isospin violation has been worked out in the $\pi$-N system to $O(p^3)$~\cite{Fettes:2001cr}
and in NN scattering to NLO~\cite{vKIV,Epelbaum:1999zn,Walzl:2000cx} but it
goes beyond the scope of the present paper to systematically include it in the three-body system.

In Section~\ref{sec:chipt} we extend the baryon $\chi$PT analysis of the
$\pi$-d scattering length to next-to-next-to leading order ($O(p^4)$). We then
formulate a modified power-counting in Section~\ref{sec:modified} with which
we derive an expression to extract the S-wave $\pi$-N scattering lengths from pionic-atom
data. A comparison with other extractions of these quantities from the
pionic-atom data is presented in Section~\ref{sec:others}. We conclude in
Section~\ref{sec:conc}.

\section{The $\pi$-d Scattering Length in Baryon $\chi$PT}
\label{sec:chipt}

We decompose the $\pi$-d scattering length
as~\cite{wein1,Beane:1998y}
\begin{equation}
a_{\pi d}=\frac{(1+\mu)}{(1+\mu /2)}(a_{\pi n} + a_{\pi
  p})+{a_{(boost)}}+{a_{(3-body)}}+\, i\,{\rm Im}~a_{\pi d} \, .
\label{eqgeneral}
\end{equation}
Each of the various contributions on the
right-hand side of Eq. (\ref{eqgeneral}) will be discussed in turn.

\subsection{The isoscalar scattering length contribution}

\begin{figure}[thbp]
   \vspace{0.5cm} \epsfysize=5.5cm
   \centerline{\epsffile{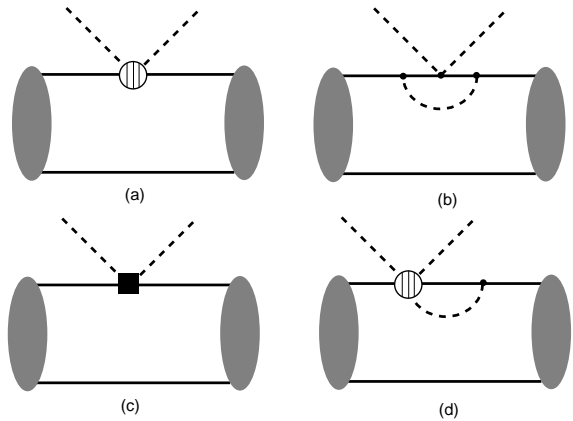}}
   \centerline{\parbox{11cm}{\caption{\label{fig1} Characteristic two-body Feynman graphs
   which contribute to the $\pi$-d scattering length at $O(p^2)$ (a), $O(p^3)$ (b) and
   $O(p^4)$ (c),(d). The dots are leading-order vertices from ${\cal L}^{(1)}_{\pi N}$, 
   the shaded blobs are from ${\cal L}^{(2,3)}_{\pi N}$, and the solid square is from
   ${\cal L}^{(4)}_{\pi N}$.}}}
\end{figure}

We first consider the ``two-body'' contributions where the pion
interacts with a single nucleon, while the other nucleon acts as a
spectator. This contribution to $a_{\pi d}$ is
proportional to the isoscalar S-wave $\pi$-N scattering
length,
\begin{equation}
a_{\pi n} + a_{\pi p}=2 a^{+}.
\end{equation}
This quantity has been computed to $O(p^3)$~\cite{Bernard:1997gq}
and $O(p^4)$~\cite{Fettes:2000xg} in baryon $\chi$PT.
Characteristic diagrams are shown in Fig.~\ref{fig1}.
The next-to-leading order (NLO= $O(p^3)$ for the amplitude) result
is~\cite{Bernard:1997gq}
\begin{eqnarray}
4\pi (1+\mu)a^+ =
\frac{M_\pi^2}{F_\pi^2} 
\left(\Delta -{{g_A^2}\over{4m}}\right)+\frac{3g_A^2{M_\pi^3}}{64\pi F_\pi^4},
\label{nloscattleven}
\end{eqnarray}
where $\Delta\equiv -4c_1+2c_2+2c_3$ and the $c_i$ are low-energy constants.
At this order there is a rather large spread in the values of $a^+$ resulting
from fitting the low-energy constants to pion-nucleon phase-shift
analyses~\cite{moj,Fettes:1998ud,Fettes:2000xg,Buttiker:2000ap}~(see Table~\ref{table-pinas}). 
As pointed out above, measurement of the width of pionic hydrogen does not ameliorate this
problem significantly, since the resulting value of $a^+$ is
consistent with zero, and the error bar is too large to facilitate
an accurate extraction of  $\Delta$.

The pionic-hydrogen measurement does constrain $a^-$ quite strongly.
On the theoretical front, we have the $O(p^4)$ expression for the
isovector S-wave $\pi$-N scattering length~\cite{Bernard:1995pa,bkm1}:
\begin{eqnarray}
4\pi (1+\mu)a^- =
\frac{M_\pi}{2F_\pi^2} +
{{4M_\pi^3}\over{F_\pi^2}}\left(\bar{D}
+{{g_A^2}\over{32 m^2}}\right) +
\frac{{M_\pi^3}}{16\pi^2 F_\pi^4}
\label{nloscattlodd}
\end{eqnarray}
where $\bar{D}\equiv\bar{d}_1 +\bar{d}_2+\bar{d}_3+2\bar{d}_5$ and
the $\bar{d}_i$ are low-energy constants that have been fit
to $\pi$-N phase shifts in Refs.~\cite{moj} and \cite{Fettes:1998ud}.
The NPE pionic-hydrogen measurement gives the value quoted in
Eq.~(\ref{eq:NPEa-}), which allows for a determination of the
combination of constants $\bar{D}$.

\subsection{The boost correction}

We now turn to the other contributions to $a_{\pi d}$ that 
can be reliably computed in baryon $\chi$PT. 
The piece ``$a_{(boost)}$'' (sometimes
known as the ``Fermi motion'' correction) is a contribution to
$a_{\pi d}$ from the graph Fig.~\ref{fig1}(a), where the sliced blob
represents a momentum-dependent vertex from the $O(p^2)$ chiral
Lagrangian, ${\cal L}^{(2)}_{\pi N}$. This vertex would vanish if
the $\pi$-N system were at rest, but it contributes to
$\pi$-d scattering, since the $\pi$-N amplitude must be boosted from
the $\pi$-N center-of-mass frame to the $\pi$-d center-of-mass frame. This
boost produces an $O(p^4)$ effect:
\begin{equation}
{a_{(boost)}}= -\frac{1}{2\pi (1+\mu /2)}
\frac{{M_\pi^2}}{4{m^3}{F_\pi^2}}({g_A^2}-8 m\, c_2 ) \left\langle
{\vec p}^{\, 2} \delta^{(3)}({\vec q}\,)\right\rangle_{\sl wf}\ ,
\label{boostcorr}
\end{equation}
where ${\vec q}\equiv {\vec p}-{\vec p^{\, \prime}}$ and
$\vec p$ and $\vec p^{\, \prime}$
are, respectively, the initial and final-state relative momenta of the
two nucleons, and 
\begin{equation}
\langle\vartheta ({\vec q})\rangle_{\sl wf}\ \equiv \ 
\int d^{\, 3}{\vec p}\ d^{\, 3}{\vec p^{\, \prime}}\ \Psi^\dagger_d ({\vec p})\
\vartheta ({\vec q})\  \Psi_d ({\vec p^{\, \prime}})\ ,
\label{sandwich}
\end{equation}
where $\Psi_d$ is the deuteron wavefunction. The three-dimensional
delta-function in Eq.~(\ref{boostcorr}) appears inside the expectation
value since this boost effect is a ``two-body'' contribution to
$a_{\pi d}$. Note that here the momentum-space deuteron wavefunctions
are normalized so that:
\begin{equation}
\left\langle \delta^{(3)}({\vec q})\right\rangle_{\sl wf}=1.
\end{equation}

\subsection{Three-body effects}

\begin{figure}[bht]
   \vspace{0.5cm} \epsfysize=3cm
   \centerline{\epsffile{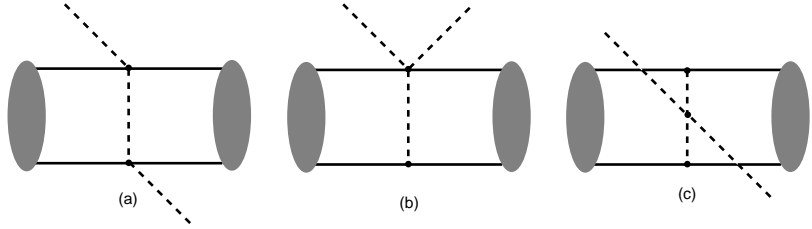}}
   \centerline{\parbox{11cm}{\caption{\label{fig2} Three-body Feynman graphs
   which contribute to the $\pi$-d scattering length at $O(p^3)$ in baryon
   $\chi$PT.}}}
\end{figure}

\noindent The dominant piece of the ``three-body'' amplitude, which occurs at
$O(p^3)$ and is shown in Fig.~\ref{fig2}, was written down some time
ago by Weinberg~\cite{wein1},
\begin{equation}
{a_{(3-body)}^{2a}}= - \frac{{M_\pi^2}}{32{\pi^4}{F_\pi^4}{(1+\mu /2)}}
\Bigg\langle\frac{1}{{\vec q\,}^{2}}\Bigg\rangle_{\sl wf}\, ;
\label{a3boda}
\end{equation}
\begin{equation}
{a_{(3-body)}^{2bc}}=\frac{{g_A^2}{M_\pi^2}}
{128{\pi^4}{F_\pi^4}{(1+\mu /2)}}
\Bigg\langle
\frac{{\vec q}\cdot{{\vec\sigma}_1}{\vec q}\cdot{{\vec\sigma}_2}}
{({\vec q\,}^{2}+{M_\pi^2})^2}\Bigg\rangle_{\sl wf}.
\label{a3bodbc}
\end{equation}

\begin{figure}[htbp]
   \vspace{0.5cm} \epsfysize=15cm
   \centerline{\epsffile{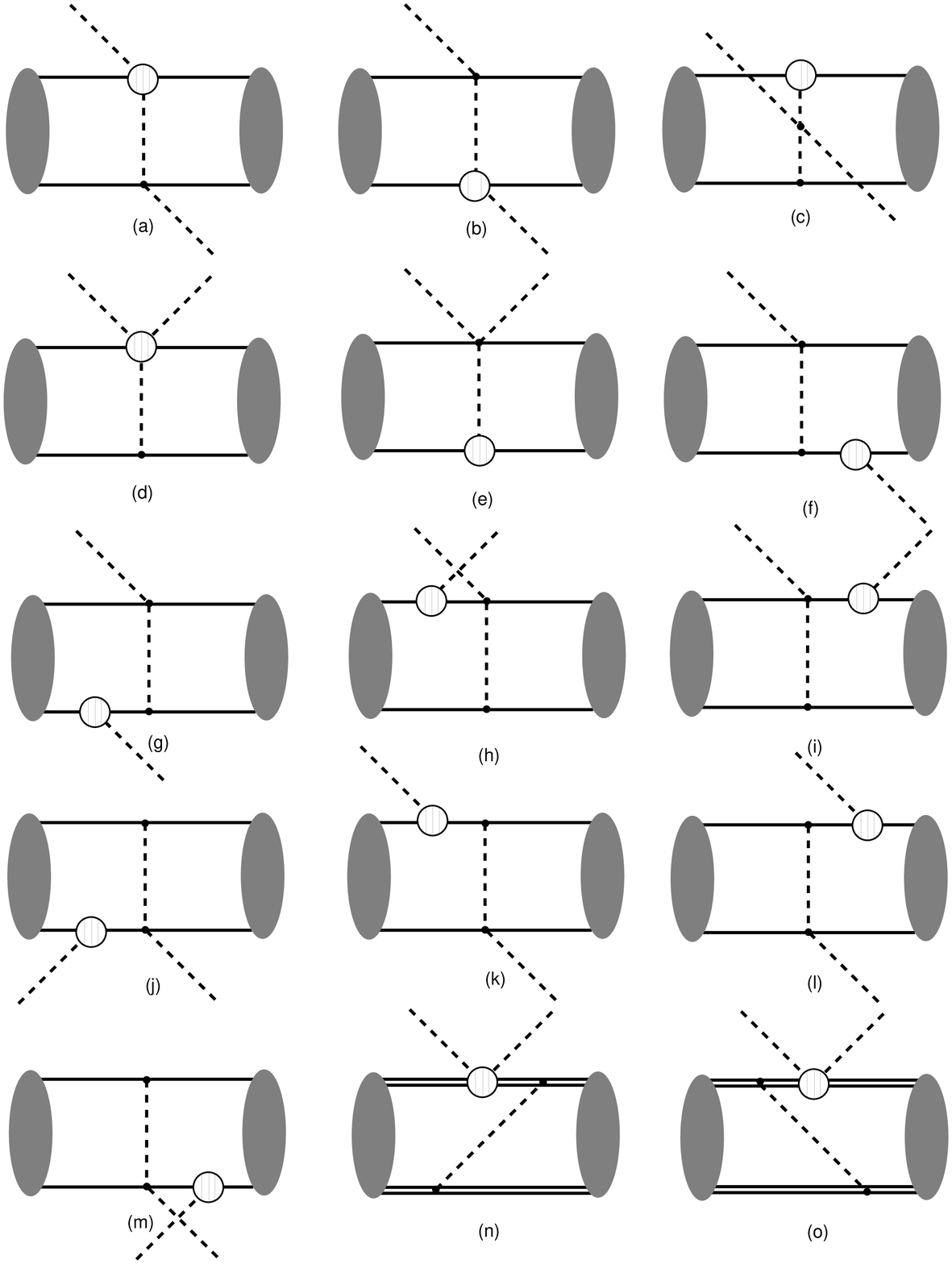}}
   \centerline{\parbox{11cm}{\caption{\label{fig3} Three-body Feynman graphs
   which contribute to the $\pi$-d scattering length at $O(p^4)$ in baryon
   $\chi$PT. The double lines indicate graphs evaluated in time-ordered
   perturbation theory.}}}
\end{figure}

Nominal $O(p^4)$ corrections to these three-body effects
are shown in Fig.~\ref{fig3}. 
The new vertices appearing in these graphs
are from ${\cal L}^{(2)}_{\pi N}$. Amusingly,
these amplitudes cancel according to the following pattern:
\begin{equation}
{\cal M}_{(c)}={\cal M}_{(d)}={\cal M}_{(e)}=0 \, ;
\end{equation}
\begin{equation}
{\cal M}_{(a)}+{\cal M}_{(b)}=0 \, ;
\end{equation}
\begin{equation}
{\cal M}_{(f)}+{\cal M}_{(g)}+{\cal M}_{(k)}+{\cal
  M}_{(l)}=0 \, ;
\end{equation}
\begin{equation}
{\cal M}_{(h)}+{\cal M}_{(j)}+{\cal M}_{(i)}+{\cal M}_{(m)}
=0 \, ;
\end{equation}
\begin{equation}
{\cal M}_{(n)}+{\cal M}_{(o)}=0 \, .
\end{equation}
{\it Thus there is no three-body correction to $a_{\pi d}$ at order
$O(p^4)$.} The sum of all corrections vanishes for a variety of
reasons, among them the threshold kinematics and the isoscalar
character of the deuteron.

\subsection{Results}

This completes the computation of $a_{\pi d}$ to $O(p^4)$ in baryon
$\chi$PT~\footnote{We relegate discussion of the imaginary part of $a_{\pi d}$
  to Section~\ref{sec:others}.}. The pieces not proportional to $a^+$ are $a_{(3-body)}$,
which is simply the $O(p^3)$ expression given in
Eqs.~(\ref{a3boda})--(\ref{a3bodbc}), and $a_{(boost)}$, which is
obtained by taking an expectation value of the one-body operator $p^2$
between deuteron wavefunctions---see Eq.~(\ref{boostcorr}).  
The deuteron wavefunction computed with the NLO ($O(p^2)$)
$\chi$PT NN potential gives a consistent determination of
$a_{\pi d}$ to $O(p^4)$. We have also shown results for the
``NNLO*'' $\chi$PT NN potential which yields a node-less deuteron 
wavefunction, and the NNLO $\chi$PT NN potential, whose deuteron 
wavefunction has two nodes. These nodes signal the presence of unphysical,
deeply-bound states. The numerical results for the various
contributions to $a_{\pi d}$ are shown in Table~\ref{tab1}. 
\renewcommand{\arraystretch}{1.2}    
\begin{table}[bht]
\begin{center}
\vskip 0.6 true cm 
\begin{tabular}{||l||c|c||c|c||c|c||}
    \hline \hline
    & \multicolumn{2}{|c||}{NLO}  &  \multicolumn{2}{|c||}{NNLO*}  &    \multicolumn{2}{|c||}{NNLO} \\ 
    \hline  
$\Lambda$ [MeV] & 500 MeV  & 600 MeV & 500 MeV & 600 MeV & 900 MeV & 1050 MeV \\
\hline  \hline
$a_{({\rm boost})}$, Fit 1   & 0.00369 & 0.00511 & 0.00375 & 0.00419 & 0.03243 & 0.03379     \\
 $a_{({\rm boost})}$, Fit 2  & 0.00350 & 0.00486 & 0.00357 & 0.00399 & 0.03084 & 0.03213  \\
 $a_{({\rm boost})}$, Fit 3  & 0.00361 & 0.00501 & 0.00368 & 0.00411 & 0.03180 & 0.03313 \\
    \hline
$a_{({\rm 3-body})}^{2a}$  & $-$0.01954  & $-$0.01864  & $-$0.02057  & $-$0.02056  & $-$0.02360   & $-$0.02257  \\
    \hline
$a_{({\rm 3-body})}^{2bc}$  & $-$0.00017 & $-$0.00035 & $-$0.00018 & $-$0.00038 &    0.00033  &    0.00018  \\
    \hline
$\langle 1/{{\vec q\,}^{2}}\rangle_{\sl wf}$\ $\lbrack M_\pi\rbrack$ & 12.6  & 12.0  & 13.2 & 13.2 & 15.2 & 14.5 \\
    \hline
$\langle 1/{|{\vec q\,}|}\rangle_{\sl wf}$\ $\lbrack M_\pi^2\rbrack$  & 8.17 & 6.20 & 10.21 & 10.20 & 26.28  & 24.78  \\
    \hline
    \hline  
\end{tabular}
\end{center}
\caption{Numerical results for the contributions to $a_{\pi d}$ given in 
Eqs.~(\ref{boostcorr}),(\ref{a3boda}) and (\ref{a3bodbc}) as well as other overlaps discussed in
the text. All scattering lengths are in units of $M_\pi^{-1}$.
$\Lambda$ is the cutoff in the NN system.}
\label{tab1}
\end{table}
The parameter set $F_\pi = 92.4$ MeV, $g_A=1.2843$ (note that we adjust $g_A$
from the Goldberger-Treiman
relation), $M_\pi = M_{\pi^+} = 139.57$ MeV and 
$m=(m_n + m_p)/2 = 938.92$ MeV is used in all expressions.
For each of the three different $\chi$PT NN potentials we vary
$\Lambda$, the cutoff in the NN system, in order to estimate our theoretical
error. The three values (Fits 1-3) quoted for $a_{({\rm boost})}$ correspond to
values of $c_2$ fit to various phase shift analyses of $\pi$-N
scattering~\cite{Fettes:1998ud} (Fits 1-3 in Table~\ref{table-pinas}).
For comparison, in Table~\ref{tab2} we give numerical results for the
same quantities computed using various ``realistic'' potential models.  
\renewcommand{\arraystretch}{1.2}    
\begin{table}[htb] 
\begin{center}
\vskip 0.6 true cm 
\begin{tabular}{||l||c|c|c|c|c||}
    \hline \hline
   & Nijm 93  & Nijm I  & Nijm II & CD Bonn 2000 & AV 18  \\
\hline  \hline
$a_{({\rm boost})}$, Fit 1  & 0.00594 & 0.00541 & 0.00623 & 0.00480 & 0.00609  \\
 $a_{({\rm boost})}$, Fit 2 & 0.00564 & 0.00515 & 0.00592 & 0.00456 & 0.00579  \\
 $a_{({\rm boost})}$, Fit 3 & 0.00582 & 0.00531 & 0.00611 & 0.00470 & 0.00597  \\
    \hline
$a_{({\rm 3-body})}^{2a}$    & $-$0.01984  & $-$0.01993  & $-$0.01954  & $-$0.02020  & $-$0.01961     \\
    \hline
$a_{({\rm 3-body})}^{2bc}$ & $-$0.00074 & $-$0.00070 & $-$0.00074 & $-$0.00055 &   $-$0.00075  \\
    \hline
$\langle 1/{{\vec q\,}^{2}}\rangle_{\sl wf}$\ $\lbrack M_\pi\rbrack$  & 12.8 & 12.8 & 12.6 & 13.0 & 12.6  \\
    \hline
$\langle 1/{|{\vec q\,}|}\rangle_{\sl wf}$\ $\lbrack M_\pi^2\rbrack$ & 7.80 & 8.26  & 7.23 & 8.70 & 7.39     \\
    \hline
    \hline  
\end{tabular}
\end{center}
\caption{Numerical results for the contributions to $a_{\pi d}$ given in 
Eqs.~(\ref{boostcorr}),(\ref{a3boda}) and (\ref{a3bodbc}) as well as other overlaps discussed in
the text, for various potential models:
Nijmegen~\cite{Nijm93}; CD-Bonn~2000~\cite{CDBonn}; AV-18~\cite{AV18}.
All scattering lengths are in units of $M_\pi^{-1}$.}
\label{tab2}
\end{table}

The following points are worth noting:
\begin{enumerate}
\item The nodes in the NNLO wavefunction cause the anomalous change
in $a_{(boost)}$ (by an order of magnitude) between the NLO and NNLO
evaluations. This anomaly can be understood as follows. The boost
correction to $a_{\pi d}$ is proportional to
$\langle p^2 \rangle_{\sl wf}$. While the sum of 
$\langle p^2 \rangle_{\sl wf}/m$ and $\langle V \rangle_{\sl wf}$---the expectation
value of the NN potential---is constrained to be $2.225$~MeV,
the NNLO potential is very deep because it supports unphysical NN states bound
by over one GeV. It naively follows that $\langle p^2 \rangle_{\sl wf}$ 
will be much larger for the NNLO wavefunction than for
wavefunctions without unphysical, deeply-bound NN states. Of course if
renormalization is done properly, this sensitivity to short-distance
physics should be absent in observables up to subleading effects in
the EFT. Given this technical issue, which is currently under 
investigation, in this paper we discard results obtained with 
the NNLO wavefunction. 

\item When evaluated with the chiral NLO and NNLO* wavefunctions the
smaller pieces of the $O(p^3)$ three-body contribution, $a_{({\rm
3-body})}^{2bc}$, are rather different from the results given in Table
1 of Ref.~\cite{Beane:1998y}. In contrast, the difference between the
evaluations of Table~\ref{tab2} and those of Ref.~\cite{Beane:1998y}
is small and arises from the different value of $g_A$ used in that
paper. The spread in $a_{({\rm 3-body})}^{2bc}$ values in Tables~\ref{tab1}
and \ref{tab2} reflects the sensitivity of this contribution to the (unphysical) short-range
behavior of the deuteron wavefunction.

\item Lastly, and most importantly, in all cases $a_{(3-body)}^{2a}$ is much
larger than any other contribution to $a_{\pi d}$---including the contribution from
$a^+$. This suggests that the baryon $\chi$PT scaling of operators is not
properly accounting for the characteristic scales present in the deuteron wavefunction.
We will return to this important point in the next section.

\end{enumerate}

\section{A Modified Power-Counting}
\label{sec:modified}

As pointed out by Weinberg~\cite{wein1}, Feynman graphs of the type
shown in Fig.~\ref{fig2}(a) are larger than their naive baryon
$\chi$PT scaling suggests because of infrared enhancements resulting
from the anomalously small deuteron binding energy. For all of the NN
wavefunctions considered in Table~\ref{tab1}, $a_{(3-body)}^{2a}$ is
larger than $a_{(3-body)}^{2bc}$ by two orders of magnitude. This is
mainly because in low-energy $\pi$-d scattering pions carry energy and
momentum that puts external pion legs, as well as internal pion legs
in diagrams such as Fig.~\ref{fig2}(a), on or near the pion
mass-shell. The amplitude in Fig.~\ref{fig2}(a) would be infrared
divergent if the deuteron were a zero-energy bound state. 
As it is, this divergence is softened by the fact that the two nucleons in the
deuteron are off their mass-shell. But the amount by which they are off their mass-shell
is far less than the pion mass, and therefore if
these effects are to be correctly accounted for in the EFT, we must
find a way to incorporate the infrared enhancement into the theory.

The completely consistent way to do this is to construct an EFT where
non-relativistic nucleons interact via contact operators, and pions
near their mass-shell are included as massless excitations.  Thus, in
$\pi$-d scattering the nucleons are largely static and the
successive scattering of a pion off the two nucleons generates a
Coulomb-like interaction between them~\cite{BeSa}. In order to have
the incoming and outgoing nucleons as well as all the pions in a
process near their respective mass-shells, there must be an equal
number of pions entering and leaving each interaction
vertex. Processes with interaction vertices that have a different number of pions
entering than leaving are represented by local operators (on the scale
of $M_\pi$) in this ``EFT with heavy pions'' (H$\pi$EFT). H$\pi$EFT is
organized as an expansion in $\gamma/M_\pi$, since physics at the
scale of the pion mass is ``integrated out'' and only appears
implicitly in the theory via constants in the Lagrangian.  Effects
contributing explicitly to $\pi$-d scattering in this EFT are two-body
$\pi$-N scattering and ``three-body'' effects in the
multiple-scattering series for the $\pi NN$ system~\cite{BeSa}. Other
effects, such as $a_{(3-body)}^{2bc}$, are non-dynamical and are represented 
by a local two-pion-four-nucleon
operator. This operator appears at leading order in the expression for
$a_{\pi d}$ and its coefficient is undetermined in H$\pi$EFT, since it 
is set by pion-scale physics. This counterterm fixes the
scale dependence of a logarithm that is formally divergent in the
infrared limit, $\gamma\rightarrow 0$.

Here we will employ an economical way of recovering the hierarchy
of scales $\gamma \ll M_\pi$ that motivates the construction of the
H$\pi$EFT. We will work in the pionful theory with a modified power
counting that accounts for the infrared enhancement of graphs in the
multiple-scattering series for the $\pi NN$ system. In practice this
means that we are able to fix the strength of the counterterm by
matching the results of H$\pi$EFT developed in Ref.~\cite{BeSa} to
those of baryon $\chi$PT derived above.

If we denote a generic momentum inside deuterium by $\vec{q}$, then
the baryon $\chi$PT power-counting is predicated upon 
$|\,{\vec q}\,|\sim M_\pi$.  One might instead expect
$|\,{\vec q}\,|\sim \sqrt{-B_d m}\equiv\gamma$, where $B_d$ and 
$\gamma$ are the deuteron binding
energy and momentum, respectively.  These two scalings would be
compatible if the deuteron binding energy were of natural
size. However, $B_d=-2.225~{\rm MeV}$ is unnaturally small on the
scale of hadronic physics, as is $\gamma=45.7025~{\rm MeV}$.  The true
scaling is thus $|\,{\vec q}\,|\sim M_\pi /3$, and
this additional $1/3$ suppression has a dramatic effect on the
ordering of operators.

\begin{figure}[bht]
   \vspace{0.5cm} \epsfysize=3cm
   \centerline{\epsffile{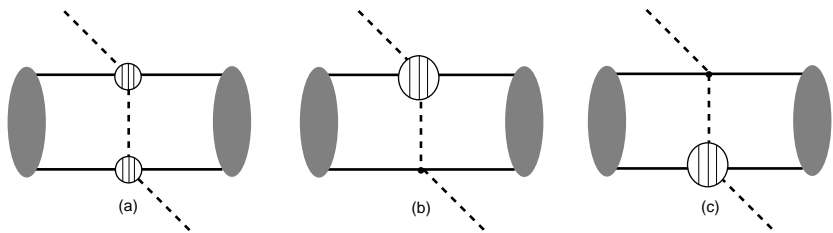}}
   \centerline{\parbox{11cm}{\caption{\label{fig5} Infrared-enhanced
   Feynman graphs which contribute to the $\pi$-d scattering length at
   $O(p^5)$ in baryon $\chi$PT and $O(Q^3)$ in the modified
   power-counting. The small shaded blobs are vertices taken from
   ${\cal L}^{(2)}_{\pi N}$ and the large shaded blobs are vertices
   taken from ${\cal L}^{(3)}_{\pi N}$ together with $O(p^3)$ one-loop
   graphs.  }}}
\end{figure}

Hence we will adopt a modified scaling of operators in which 
$|\,{\vec q}\,|\sim p^2$ where $p$ is the baryon
$\chi$PT expansion parameter~\footnote{Yet another power-counting for
$\pi$-d scattering in which pions are purely perturbative has been
studied in Ref.~\cite{Borasoy:2001gq}. Unfortunately, it is now known
that pions are strongly non-perturbative in the deuteron
channel~\cite{Fleming:1999ee}.}.  Amplitudes involving $M_\pi$ in the
denominator are now suppressed, just as is the case in H$\pi$EFT. The expansion
parameter for this modified power-counting is denoted by $Q$, where
$Q$ is the same size as $p$ in baryon $\chi$PT but, of course,
leads to a different ordering of operators.

It is easy to check that in the modified power-counting
$a_{(3-body)}^{2a}$ is $O(Q)$ while $a_{(3-body)}^{2bc}$ is
$O(Q^5)$; thus, on the basis of scaling, $a_{(3-body)}^{2bc}$ is expected 
to be much smaller than $a_{(3-body)}^{2a}$. Meanwhile, the boost correction of
Eq.~(\ref{boostcorr}) involves the expectation value inside the
deuteron of two powers of momentum and a three-dimensional
delta-function; and so scales as $O(Q^3)$.

In Table~\ref{tab3} we exhibit various contributions for a
characteristic wavefunction and their scalings in both baryon $\chi$PT
and in the modified power-counting. The modified counting does a much
better job of predicting the sizes of various mechanisms, presumably
because it includes the full hierarchy of scales 
$\gamma \ll M_\pi \ll \Lambda_\chi$, where $\Lambda_\chi$ sets the breakdown 
scale of the chiral expansion.
\vspace{-1cm}
\renewcommand{\arraystretch}{1.2}    
\begin{table}[htb] 
\begin{center}
\vskip 0.6 true cm 
\begin{tabular}{||l||c|c|c|c||}
    \hline \hline
   & Baryon $\chi$PT  & Modified PC & NLO (500 MeV) & NLO (600 MeV)  \\
\hline  \hline
$a_{({\rm 3-body})}^{2a}$    & $O(p^3)$ & $O(Q^1)$  & $-$0.01954   &
$-$0.01864  \\
    \hline
$a_{({\rm boost})}$ (Fit 1) & $O(p^4)$ & $O(Q^3)$  & $\;\;\;$0.00369 &
$\;\;\;$0.00511  \\
    \hline
$a_{({\rm 3-body})}^{5}$  & $O(p^5)$ & $O(Q^4)$  & $\;\;\;$0.00184   &
$\;\;\;$0.00175  \\
    \hline
$a_{({\rm 3-body})}^{2bc}$  & $O(p^3)$ & $O(Q^5)$  & $-$0.00017  &
$-$0.00035   \\
    \hline
$a_{({\rm 3-body})}^{range}$, Ref.~\cite{Norbert} & $O(p^5)$ & $O(Q^5)$  
& $\;\;\;$0.00120   & $\;\;$0.00041  \\
    \hline
    \hline  
\end{tabular}
\end{center}
\caption{Scaling of various contributions to the scattering length in baryon
  $\chi$PT and in the modified power-counting as discussed in the text. All
  matrix elements are evaluated with both the NLO ($500$~MeV) 
and the NLO ($600$~MeV) wavefunctions.}
\label{tab3}
\end{table}

In the modified power-counting we do not yet have the full result
at $O(Q^4)$. In particular, graphs of the type shown in
Fig.~\ref{fig5} are promoted from their $O(p^5)$ scaling in baryon
$\chi$PT to $O(Q^3)$ in the modified counting. 
These graphs are corrections to
$a_{(3-body)}^{2a}$ with an insertion from ${\cal L}^{(2)}_{\pi N}$ at each
vertex (Fig.~\ref{fig5}(a)), or
an insertion from ${\cal L}^{(1)}_{\pi N}$ at one vertex and
${\cal L}^{(3)}_{\pi N}$ at the other vertex (Fig.~\ref{fig5}(b,c)).
Computation of these graphs is straightforward. 
Fig.~\ref{fig5}(a) evaluates to
\begin{equation}
{a^{4a}_{(3-body)}}=  \frac{M_\pi^4}{16{\pi^4}{F_\pi^4}{(1+\mu /2)}}
\,\left(\Delta - {{g_A^2}\over{4m}}\right)^2\,
\Bigg\langle
\frac{1}{{\vec q\,}^{2}}
\Bigg\rangle_{\sl wf}\, ,
\label{4a}
\end{equation}
while Fig.~\ref{fig5}(b) and Fig.~\ref{fig5}(c) are given by
\begin{equation}
{a^{4bc}_{(3-body)}}= - \frac{M_\pi^4}{8{\pi^4}{F_\pi^4}{(1+\mu /2)}}
\,\Bigg\lbrack
4\left(\bar{D}+{{g_A^2}\over{32m^2}}\right) +\frac{1}{16\pi^2 F_\pi^2}
\Bigg\rbrack\,
\Bigg\langle
\frac{1}{{\vec q\,}^{2}}
\Bigg\rangle_{\sl wf}\,  .
\label{4bc}
\end{equation}

The complete $O(Q^3)$ expression for the real part of the $\pi$-d scattering length
is
\begin{equation}
{\rm Re}\,a_{\pi d}=2\frac{(1+\mu)}{(1+\mu /2)}a^+
+{a_{(boost)}}+{a^{2a}_{(3-body)}}+{a^{4a}_{(3-body)}}+{a^{4bc}_{(3-body)}}
\label{apidinmodpc}
\end{equation}
where the various terms can be found in Eqs.~(\ref{nloscattleven}), (\ref{boostcorr}), (\ref{a3boda}),
(\ref{4a}) and (\ref{4bc}), respectively. The astute reader will have noticed that to the order
we are working we can rewrite Eq.~(\ref{apidinmodpc}) in the form
\begin{equation}
{\rm Re}\,a_{\pi d}=2\frac{(1+\mu)}{(1+\mu /2)}\,\left( a^+ \, +\,
(1+\mu )\Big\lbrack (a^+)^2-2(a^-)^2 \Big\rbrack 
\frac{1}{2{\pi^2}}\Bigg\langle 
\frac{1}{{\vec q\,}^{\, 2}}
\Bigg\rangle_{\sl wf}\,\right) \,+\, {a_{(boost)}}\, +\, O(Q^4).
\label{MOD2}
\end{equation}
Thus, to $O(Q^3)$ the EFT with modified power-counting reproduces the simple intuitive picture
of a single-scattering term plus a pion-exchange correction~\cite{eric}.

Notice that in the amplitudes of Fig.~\ref{fig5},
the momentum dependence of the vertices from ${\cal
L}^{(2)}_{\pi N}$ and ${\cal L}^{(3)}_{\pi N}$ has been
neglected. While these corrections occur at the same
order in baryon $\chi$PT as Eq.~(\ref{4a}) and Eq.~(\ref{4bc}) ($O(p^5)$), 
in the modified power-counting the momentum-dependent corrections
begin at $O(Q^5)$ and are thus suppressed by two orders. 
These ``range'' corrections were recently computed in baryon $\chi$PT~\cite{Norbert}. 
The size of these contributions is consistent with the $O(Q^5)$
scaling (see Table~\ref{tab3}). There are, of course, many other 
contributions at $O(Q^5)$, including two-pion-four-nucleon local operators with 
unknown coefficients~\footnote{It is, in a sense, unfortunate that these
  effects are suppressed, since one of the local operators at $O(Q^5)$ encodes information
  about the leading quark-mass dependence of the deuteron binding energy~\cite{Beane:2001bc,BeSa2}.}.
As we will see below, there are many more important contributions at $O(Q^4)$ 
which are easily computed in the EFT. Hence we will not consider any $O(Q^5)$
effects in isolation, but will instead perform the complete
calculation up to $O(Q^4)$ in the modified power-counting.

\begin{figure}[ht]
   \vspace{0.5cm} \epsfysize=2.3cm
   \centerline{\epsffile{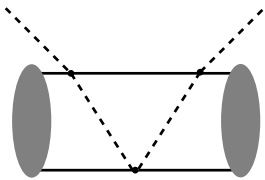}}
   \centerline{\parbox{11cm}{\caption{\label{fig4} One-loop three-body
    diagram that contributes at $O(Q^4)$ in the modified power-counting and
    at $O(p^5)$ in baryon $\chi$PT.}}}
\end{figure}

It is surprisingly easy to compute the $O(Q^4)$ corrections to
the $O(Q^3)$ result, Eq.~(\ref{MOD2}). There are no further two-body
effects from the $\pi$-N amplitude at $O(Q^4)$. There are, of course,
diagrams similar to those in Fig.~\ref{fig5} and Fig.~\ref{fig2}(a),
but with higher-order momentum-independent vertices. However, these
corrections are already included in Eq.~(\ref{MOD2}) since they are
subsumed in $a^\pm$. There is then a single three-body one-loop graph,
Fig.~\ref{fig4}, representing the next term in the $\pi NN$
multiple-scattering series. This contributes at
$O(Q^4)$. Fig.~\ref{fig4} is one of many $O(p^5)$ baryon $\chi$PT
graphs that contribute to $a_{\pi d}$.  However, the other graphs all have
intermediate pions that are far from their mass-shell, and so are
of higher order in the modified power-counting.

We have evaluated Fig.~\ref{fig4} using the leading-order $\chi$PT $\pi$-N
vertex for $\pi$-N scattering and have neglected contributions to the
loop integral where pions are $\sim M_\pi$ off their mass-shell. These neglected
contributions are again higher order in the modified power-counting. This
gives:
\begin{eqnarray}
{a_{(3-body)}^5}=  \frac{1}{64{\pi^4}{(1+\mu /2)}}
\left(\frac{M_\pi}{2 F_\pi^2}\right)^3 \Bigg\langle
 \frac{1}{|{\vec q\,}|} \Bigg\rangle_{\sl wf}\ .
\label{atriangle}
\end{eqnarray}
Notice that, as in Eq.~(\ref{a3boda}), this matrix element has a Coulomb-like
propagator, which signals the presence of an infrared logarithm. It is this
enhancement that is captured by the modified power-counting.

Our final formula for the $\pi$-d scattering length, valid to $O(Q^4)$ in the
modified power-counting, is
\begin{eqnarray}
{\rm Re}\,a_{\pi d}&=&2\frac{(1+\mu)}{(1+\mu /2)}\,\left( a^+ \, +\, 
(1+\mu )\Big\lbrack (a^+)^2-2(a^-)^2 \Big\rbrack 
\frac{1}{2{\pi^2}}\Bigg\langle 
\frac{1}{{\vec q\,}^{\, 2}}\Bigg\rangle_{\sl wf}\qquad\qquad\qquad\qquad \right. \nonumber\\
&& \left.  \qquad\qquad\qquad\qquad\quad
+(1+\mu )^2\Big\lbrack (a^+)^3-2(a^-)^2(a^+-a^-) \Big\rbrack 
\frac{1}{4{\pi}}\Bigg\langle 
\frac{1}{|{\vec q}\, |}\Bigg\rangle_{\sl wf}\, \right) \nonumber\\
&& \qquad\qquad\qquad\qquad\qquad\quad
\,+\, {a_{(boost)}}\, +\, O(Q^5)\ .
\label{thefullthing}
\end{eqnarray}
where we have again subsumed higher-order effects into the 
$\pi$-N scattering lengths. Notice that since local two-pion-four-nucleon
operators are not enhanced in the modified power-counting they appear
at the same order as in baryon $\chi$PT, namely fifth order, and do
not affect the fourth-order result given in Eq.~(\ref{thefullthing}).

\begin{figure}[ht]
   \vspace{0.5cm} \epsfysize=7.5cm
   \centerline{\epsffile{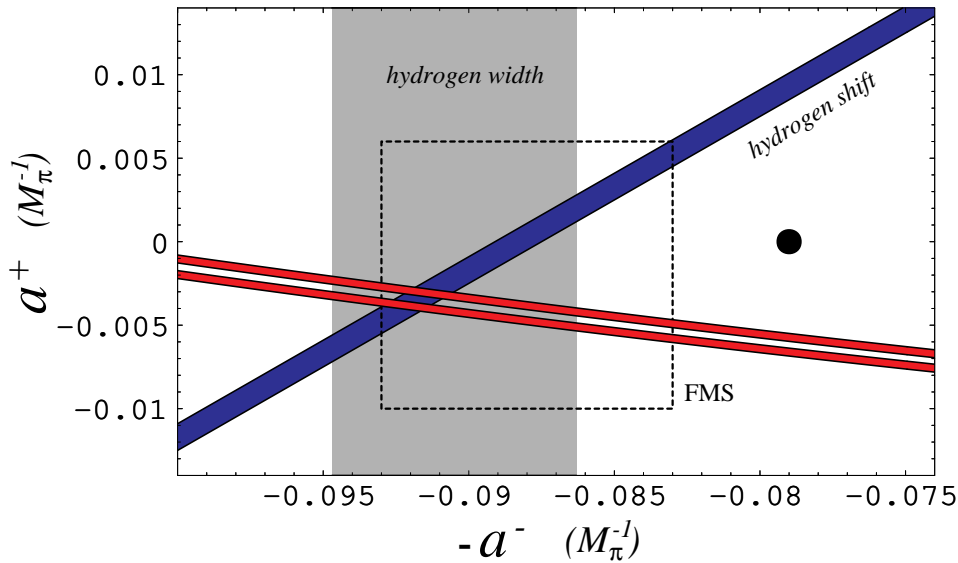}}
   \centerline{\parbox{11cm}{\caption{\label{figschro} Plot of $a^+$ 
    vs $-a^-$. The light shaded region and the dark band are from the experimental
    pionic-hydrogen width and shift, respectively, taken from Ref.~\cite{Schroder:uq}.
    The dotted line encompasses the constraints from $\pi$-N phase shift data
    and is taken from Ref.~\cite{Fettes:1998ud}.
    The dot is leading order $\chi$PT (current algebra). The two parallel bands are from
    Eq.~(\ref{thefullthing}) evaluated with the NLO wavefunction with an
    ultraviolet cutoff of $500$~MeV (upper curve) and $600$~MeV (lower curve).}}}
\end{figure}

Equation~(\ref{thefullthing}) is easily inverted to give a constraint on
$a^+$ and $a^-$ in terms of the experimentally-determined $a_{\pi d}$
from Eq.~(\ref{eq:pionicdeut}). In order to evaluate the constraint we
need the matrix elements $\langle 1/{\vec q}^{\, 2}\rangle_{\sl
wf}$ and $\langle 1/|{\vec q}\,| \rangle_{\sl wf}$, as well as
$a_{(boost)}$ (see Tables~\ref{tab1} and \ref{tab2}). The boost effect does 
depend on the chiral low-energy
constant, $c_2$, but this counterterm was fixed in
Ref.~\cite{Fettes:1998ud} by fitting $\pi$-N phase shift data. The
resulting relationship between $a^-$ and $a^+$ is displayed in
Fig.~\ref{figschro}. Notice that we choose to plot $a^+$ vs $-a^-$, in order 
to facilitate comparison with other extractions.

The two bands in Fig.~\ref{figschro} correspond to two different
evaluations of the deuteron matrix elements involved: one with the NLO
wavefunction with an ultraviolet cutoff of $500$~MeV (upper curve) and
the other with a cutoff of $600$~MeV (lower curve).  In each case the
finite width of the band arises from the error associated with the
pionic-deuterium measurement of $a_{\pi d}$. 
The spread in the values of $c_2$ arising from Fits 1-3 of Table~\ref{table-pinas}
also falls within these two bands. The result for $a^+$
found using ``realistic'' NN wavefunctions falls between these two
bands, as, in most cases, would the effect of a typical $O(Q^5)$ 
correction. The experimental curves for the pionic-hydrogen shift
and width are taken from Ref~\cite{Schroder:uq}, while the bounds from
baryon $\chi$PT fits to $\pi$-N phase shift data are taken from
Ref~\cite{Fettes:1998ud}.  From the overlap region in this plot we
extract the S-wave $\pi$-N scattering lengths. We find 
\begin{eqnarray}
a^{-}&=&[0.0918 (\pm 0.0013)\;] {M_\pi^{-1}}\ ; \nonumber\\
a^{+}&=&[-0.0034 (\pm 0.0007)\;] {M_\pi^{-1}}\ .
\label{finalisoscalarandvector}
\end{eqnarray}
The theoretical error bars should be considered somewhat optimistic
given our neglect of isospin violation.

\section{Comparison with other extractions}
\label{sec:others}

The results of Eq.~(\ref{finalisoscalarandvector}) are consistent with other
recent extractions that do not make use of the EFT
framework~\cite{baru,Ericson:2000md}. In particular, they are
in agreement with very recent work of Ericson and
collaborators~\cite{Ericson:2000md} who find 
$a^{-}=[0.0900 (\pm 0.0016)\;] {M_\pi^{-1}}$ and 
$a^{+}=[-0.0017 (\pm 0.0010)\;]{M_\pi^{-1}}$. 
The calculation of Ref.~\cite{Ericson:2000md}
includes the basic mechanisms present in the EFT employed here, but also includes
additional contributions to $a_{\pi d}$ which are higher order in the modified
power-counting.

For instance, the analysis of Ref.~\cite{Ericson:2000md} includes a 
``dispersive contribution'' (with a sizeable error bar) that is numerically comparable to our
$O(Q^3)$ boost correction (but with opposite sign). This contribution accounts
for most of the numerical difference between the two analyses. However, we find that the
``dispersive contribution'' occurs only beyond the order to which we have worked in the
modified power-counting. 
\begin{figure}[h]
   \vspace{0.5cm} \epsfysize=3cm
   \centerline{\epsffile{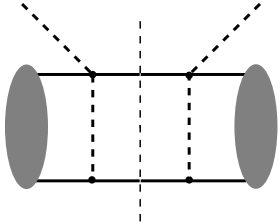}}
   \centerline{\parbox{11cm}{\caption{\label{fig6} Characteristic
   one-loop three-body diagram that contributes at $O(p^5)$ in
   $\chi$PT. This graph gives a contribution to the imaginary
   part of the $\pi$-d scattering length. }}}
\end{figure}

One graph which contributes to both the dispersive and absorptive parts
of $a_{\pi d}$ is shown in Fig.~\ref{fig6}. This three-body one-loop
graph is one of the leading contributions to ${\rm Im}~a_{\pi d}$.
It is $O(p^5)$ in baryon $\chi$PT and $O(Q^6)$ in the modified power 
counting~\footnote{Since pion number is not conserved at the interaction vertices, this
graph appears as a two-pion-four-nucleon local operator in H$\pi$EFT.}. 
The $O(Q^6)$ scaling of Fig.~\ref{fig6} is seemingly at odds with
the experimental value of ${\rm Im}~a_{\pi d}$ given in
Eq.~(\ref{eq:pionicdeut}), which is consistent with an
$O(Q^3)$ effect. It is easy to see why the scaling might break down
for the imaginary part of the scattering length.
The absorptive part of Fig.~\ref{fig6} is related by unitarity to the 
$\pi$ d$\rightarrow N N$ process at threshold:
\begin{equation}
{\rm Im}\, a_{\pi d}=\frac{1}{4\pi}\lim_{q\rightarrow 0} q\, \sigma
(\pi^- d \rightarrow n n)
\end{equation}
where $q$ is the pion three-momentum in the $\pi$-d center-of-mass
system. A study of $\pi$ d$\rightarrow N N$ has been made in
Ref.~\cite{daRocha:1999dm}.  Unfortunately baryon $\chi$PT applied to
pion production significantly underpredicts the data. Presumably this
is due to the large expansion parameter inherent in the
pion-production process: since in this diagram momenta of order
$\sqrt{m M_\pi}$ are present, the baryon $\chi$PT expansion parameter
is $\sqrt{p}$ rather than $p$.  Clearly this problem is present in any
attempt to compute the absorptive contribution to the $\pi$-d
scattering length in EFT, and is beyond the scope of this paper.
Whether this is also an issue for the dispersive part is worth investigating,
given the sizeable contribution claimed in Ref.~\cite{Ericson:2000md}.
We should also mention that Refs.~\cite{baru} and \cite{Ericson:2000md} give an estimate
of isospin violation that is consistent with an $O(Q^5)$ effect in the modified
power counting. A systematic investigation of this very important issue within EFT is needed.
Finally, our result for the isoscalar scattering
length is consistent with a previous baryon $\chi$PT determination at
$O(p^3)$~\cite{Beane:1998y}. Because a hybrid approach employing
phenomenological wavefunctions was used in that paper,
no theoretical error bar was given.

\section{Conclusion}
\label{sec:conc}

In this paper we have developed an EFT for the $\pi$-d scattering
length. While naively, baryon $\chi$PT, as modified by
Weinberg~\cite{Weinberg:1990rz} for the NN system, captures
the correct hierarchy of scales, we have seen that there are several
subtleties in the analysis of $\pi$-d scattering near threshold, which
suggest a more practical power-counting. The new ingredient is the
observation that one may develop a hierarchy between the pion mass,
$M_\pi$ and the deuteron binding momentum, $\gamma$. One can then pick
out the pieces of the amplitude that become large in the formal limit
in which $\gamma$ goes to zero and $M_\pi$ is held
fixed~\cite{BeSa}. This does not, of course, indicate that anything is
amiss with baryon $\chi$PT power-counting. Baryon $\chi$PT should work
just fine for $\pi$-d scattering in the threshold region, but 
high-order calculations are required if truly accurate results are desired.
The advantage of the modified power-counting, and of EFT generally, is
that it immediately isolates the large contributions, without
requiring explicit calculations of matrix elements that end up being
smaller than the theoretical error.

\smallskip

\noindent{\bf Acknowledgments:}

\smallskip

\noindent 
We thank Martin Savage and Iraj Afnan for valuable conversations.  DRP
thanks Flinders University, where part of this work was performed, for
its hospitality.  This research was supported in part by the
U. S. Department of Energy under grants DE-FG03-97ER41014 (SRB) and
DE-FG02-93ER40756 (DRP), and by Deutsche Forschungsgemeinschaft under
grant GL 87/34-1.
\newpage%

\newpage


\end{document}